# Three-orbital study on the orbital distillation effect in the high $T_c$ cuprates

H. Sakakibara[a*], K. Suzuki[b], H. Usui[a], K. Kuroki[a], R. Arita[b,e], D.J. Scalapino[c] and H. Aoki[d]

[a]*Department of Engineering Science, The University of Electro-Communications, Chofu, Tokyo 182-8585, Japan*
[b]*Department of Applied Physics, The University of Tokyo, Hongo, Tokyo 113-8656, Japan*
[c]*Physics Department, University of California, Santa Barbara, California 93106-9530, USA*
[d]*Department of Physics, The University of Tokyo, Hongo, Tokyo 113-0033, Japan*
[e]*JST, PRESTO, Kawaguchi, Saitama 332-0012, Japan*

**Abstract**

Our recent study has revealed that the mixture of the $d_{z^2}$ orbital component into the Fermi surface suppresses $T_c$ in the cuprates such as La$_2$CuO$_4$. We have also shown that applying hydrostatic pressure enhances $T_c$ due to smaller mixing of the Cu4$s$ component. We call these the "orbital distillation" effect. In our previous study, the 4$s$ orbital was taken into account through the hoppings in the $d_{x^2-y^2}$ sector, but here we consider a model in which of the $d_{x^2-y^2}$, $d_{z^2}$ and 4$s$ orbitals are all considered explicitly. The present study reinforces our conclusion that smaller 4$s$ hybridization further enhances $T_c$.

## 1. Introduction

One of the important, and still not fully understood, problems associated with the high-$T_c$ cuprates is how to optimize their superconducting transition temperature, $T_c$ [1,2]. It is well known that $T_c$ varies strongly with the number of CuO$_2$ layers. However, even within the single layered cuprates, it is also known that there is significant material dependence of $T_c$, e.g., La$_2$CuO$_4$ (La214; $T_c$ ~40K) and HgBa$_2$CuO$_4$ (Hg1201; $T_c$ ~100K)[3]. In La214, the shape of the Fermi surface is observed to be square compared to that of Hg1201, so a warped Fermi surface apparently favours superconductivity [4,5]. This, however, conflict with theoretical many-body studies of Hubbard-type models that indicate warped Fermi surfaces are unfavourable for superconductivity. This has remained a long-standing puzzle in the field of the study of the cuprates [6].

As for the shape of the Fermi surface, some studies have pointed out that the contribution of the $d_{z^2}$ [7,8,9] orbital and hence the apical oxygen height($h_O$) are important[10,11,12,13,14,15,16,17,18]. On the other hand, some theoretical models which include the effect of the $d_{z^2}$ orbital have explained the material dependence of $T_c$ [19,20,21,22]. However, there seems to have been no persuasive solution for this problem between $T_c$ and the shape of Fermi surface, at least, within many-body approaches for the Hubbard-type models with realistic values of the on-site $U$ [23].

To solve this puzzle, we have introduced a $d_{x^2-y^2}$-$d_{z^2}$ two-orbital model in which not only the conventionally-considered $d_{x^2-y^2}$ Wannier orbital but also the $d_{z^2}$ orbital is explicitly considered. By applying many-body analysis to the two-orbital model, we have shown that the admixture of the $d_{z^2}$ orbital in the Fermi surface Bloch states is crucial in understanding the material dependence of $T_c$ in the cuprates, i.e., in relatively low $T_c$ materials such as La214, the $d_{z^2}$ orbital suppresses the warping of the Fermi surface which would enhance $T_c$. However the increase of the $d_{z^2}$ component of the density of states to the Fermi surface overcomes this enhancement and suppresses $T_c$. As a result, La214 with a relatively squared Fermi surface nevertheless exhibits a low $T_c$ [17,18].

More recently, we have found in a study of the hydrostatic pressure enhancement of $T_c$ [23,24] that the Cu4$s$ orbital significantly affects $T_c$ in cuprates in which the $d_{z^2}$ orbital mixture is small[25]. In that study, the contribution from the 4$s$ orbital was implicitly included in the hoppings of the 3$d_{x^2-y^2}$ and the 3$d_{z^2}$ orbitals, namely, a smaller 4$s$ contribution reduces the second and third nearest neighbour hoppings between 3$d_{x^2-y^2}$ Wannier orbitals (as will be explained later in detail), thereby reducing the warping of the Fermi surface and enhancing $T_c$. Combining the effect of 3$d_{z^2}$ and 4$s$, we have concluded that the "orbital distillation" effect enhances $T_c$ in the cuprates.

In the present work, we consider the 4$s$ orbital explicitly in the model Hamiltonian, which now contains all of the 3$d_{x^2-y^2}$, 3$d_{z^2}$ and 4$s$ orbitals, to study the effect of orbital distillation on superconductivity. Our ultimate goal along this line of study is to consider the possibility of new materials which can have $T_c$ even higher than the cuprates.

*Corresponding author. Tel.: +81-42-443-5559; fax: +81-42-443-5563.
E-mail address: hiro_rebirth@vivace.e-one.uec.ac.jp.

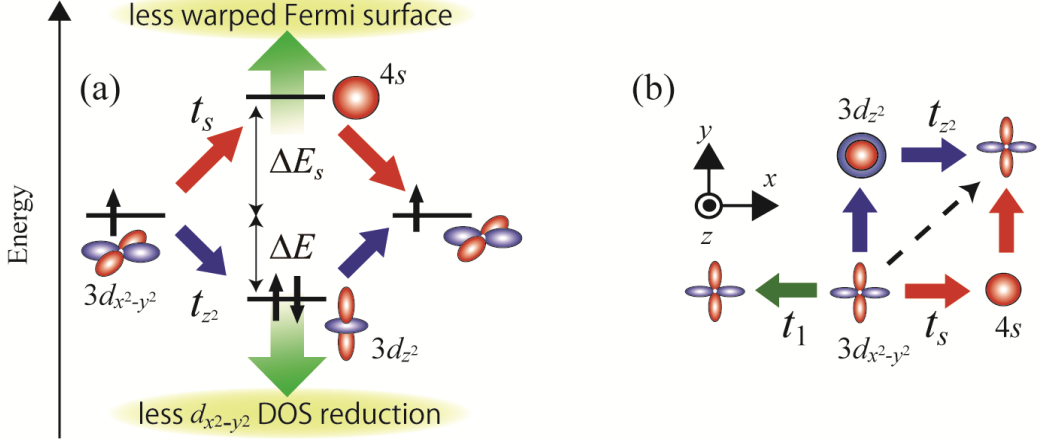

Fig.1 (a) Definitions of $\Delta E_s$, $t_s$, $\Delta E$ and $t_{z^2}$. The thickest arrows represent the orbital distillation (see text). Two paths via Cu4$s$ or Cu3$d_{z^2}$ (solid arrows $t_s$ and $t_{z^2}$) which give the effective diagonal hopping (dashed arrow).

## 2. Calculation methods

First, we perform first-principles band calculation to obtain the structural parameter of HgBa$_2$CuO$_4$ [26]. Namely, we calculate the total energy varying the lattice constants, and fit the result with the standard Burch-Marnaghan formula [27] to obtain the structure at the most stable point. From this we obtain the crystal structure, which turns out to be within 1% discrepancy from the experimentally determined lattice constants [28]. In the Hg compound, the effect of the 3$d_{z^2}$ orbital is negligible because of the large level offset $\Delta E$ between the 3$d_{x^2-y^2}$ and the 3$d_{z^2}$ orbitals, so that we can focus on the effect of the 4$s$ orbital. From this we construct maximally localized Wannier orbitals [29,30] to obtain the hopping integrals of the present three-orbital model, in which we consider the 3$d_{x^2-y^2}$ orbital, the 3$d_{z^2}$ orbital and the 4$s$ orbital explicitly as discussed above.

### 1.1. Applying the fluctuation exchange approximation

The electron-electron interactions considered in the present study are the following: the on-site intra-orbital Coulomb repulsion $U$, the inter-orbital repulsion $U$', the Hund's coupling $J$ and pair-hopping $J$'. Here we observe the orbital SU(2) requirement, $U-U$'$=2J$. Here we fix the values at $U$ =3.0 eV, $U$ '=2.4 eV and $J$ =$J$ '=0.3 eV. In the recent estimations by first-principles, $U$ in cuprates is considered to be 7-10$t$ (namely, about 3-4.5 eV) and $J$ ($J$')~0.1$U$, so the values chosen here are within the widely accepted range.

Then we apply the fluctuation exchange approximation (FLEX)[31,32,33] to the present model to obtain the Green's function renormalized by the many-body self-energy correction. In FLEX, the contributions from bubble and ladder diagrams are included in the self-energy, for which we solve the Dyson's equation in a self-consistent manner. Then we substitute the Green's function to linearized Eliashberg equation to evaluate the strength of the superconducting instability. The eigenvalue λ of Eliashberg equation reaches unity at $T_c$, so we can use λ at a fixed temperature as a measure of $T_c$. We set $T$=0.015eV, and the number of electrons per copper site to be $n$=2.85 (i.e., 15% doped in the main band). We take the 32×32×4 $k$-point meshes and 1024 Matsubara frequencies.

## 3. Results and discussion

### 3.1 Effects of the Cu4s orbital

Due to the symmetry of the wavefunctions, the Cu4$s$ orbital can mediate the electron hopping path 3$d_{x^2-y^2}$ →4$s$→3$d_{x^2-y^2}$, so that the 4$s$ orbital effectively enhances the second and third neighbour hoppings between 3$d_{x^2-y^2}$ orbitals (Fig.1) [11]. In our previous two-orbital model, the 4$s$ orbital effect is implicitly included in the 3$d_{x^2-y^2}$ and the 3$d_{z^2}$ Wannier orbitals, where the 4$s$ effect is taken into account mainly via the second ($t_2$) and third ($t_3$) neighbour hoppings of the 3$d_{x^2-y^2}$ Wannier orbital sector, and hence the warping of the Fermi surface [17].

In the present three orbital model, the warping of the Fermi surface is controlled by two parameters; the level offset $\Delta E_s$ and the nearest neighbour inter-orbital hopping $t_s$ between the 3$d_{x^2-y^2}$ and 4$s$ orbitals. In the following, we vary these two parameters hypothetically to see more directly the basic mechanism of $T_c$ suppression by the 4$s$ orbital.

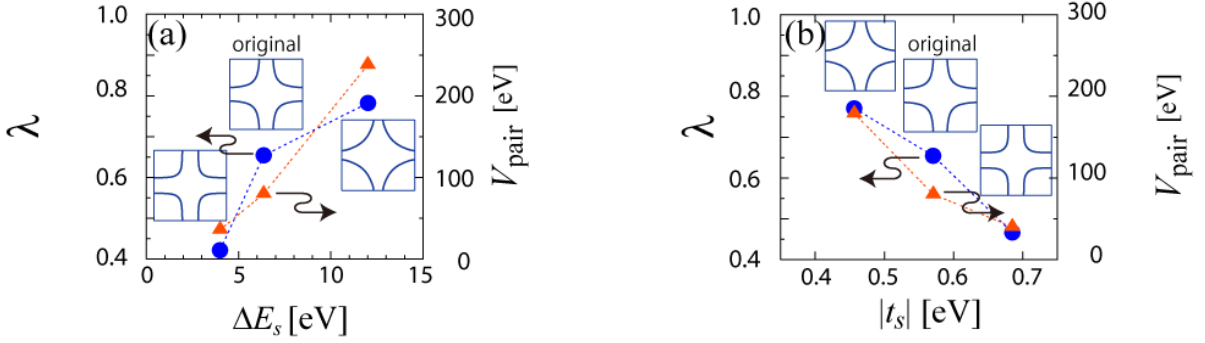

Fig.2 Eigenvalue λ(circle) and the pairing interaction $V_{pair}$(triangle) plotted against $\Delta E_s$(with a fixed $|t_s|$=0.57 eV ; left panel) or $|t_s|$( with a fixed $\Delta E_s$ =6.4 eV; right) for $HgBa_2CuO_4$.

*3.2 Effects of $\Delta E_s$ and $t_s$*

First, we focus on the effect of $\Delta E_s$. In the panel (a) of Fig.2, we plot the relationship between λ and $\Delta E_s$. We can see that λ increases with $\Delta E_s$. To explore the mechanism for the enhanced $T_c$, we have also plotted the maximum value of the pairing interaction $V_{pair}$ along with the shape of Fermi surface at $k_z$=0 in Fig.2. $V_{pair}$ is seen to increase as the warping of the Fermi surface is reduced, i.e., $V_{pair}$ is enhanced by the nesting of the Fermi surface. In other words, the reduction of the 4*s* orbital mixing improves the Fermi surface nesting and this enhances the spin-fluctuation mediated superconductivity [6].

As for the effect of the hopping $t_s$, we obtain a similar result (panel (b) in Fig.2), namely the reduction of $|t_s|$ makes the Fermi surface less warped, hence higher $T_c$. From this viewpoint, *searching for materials with reduced $t_s$* may be a promising way to obtain high $T_c$ materials. Here let us make a comment on the pressure effect studied in our previous paper [25]. Both $t_s$ and $\Delta E_s$ increase with applying hydrostatic pressure, but the effect of the increased $t_s$ is nearly cancelled out by the increase of the nearest neighbour hopping $t_1$ (note that the warping of the Fermi surface is determined by the *ratio* of the nearest and distant hoppings), so that only the effect of the $\Delta E_s$ increase remains, which enhances (suppresses) the $T_c$ (warping of Fermi surface).

*3.3 Orbital distillation*

In some cuprates such as La214, the bending of the Fermi surface induced by the 4*s* orbital is cancelled by the $3d_{z^2}$ orbital because $\Delta E$ and $\Delta E_s$ have opposite signs (see Fig.1 (a)). Namely, the 4*s* orbital mediates a diagonal hopping whose sign is opposite to the one mediated by the $3d_{z^2}$ orbital [17,18]. However, for small $\Delta E$, the $3d_{z^2}$ orbital component partially replaces the $3d_{x^2-y^2}$ orbital component, and this reduction of the density of $3d_{x^2-y^2}$ states overcomes the $T_c$ enhancement associated with the reduced warping of the Fermi surface. The reason why the $3d_{z^2}$ orbital cannot be integrated out before the many-body analysis (in contrast to the case of 4*s* orbital) comes from the fact that $\Delta E$ (~1eV) is considerably smaller than $U$ (~3eV), while $\Delta E_s$(~7eV) is much larger [18].

Note that there are two effects that can enhance $T_c$ when the 4*s* contribution is reduced : (i)the reduction of the warping due to less $3d_{x^2-y^2}$-4*s* hybridization, and (ii)the increase of the $d_{x^2-y^2}$ density of states (the same effect as in the increase of $\Delta E$). Actually, the 4*s* density of states in the Fermi surface is small (compared to that of $d_{z^2}$ in La214), so that effect (i) governs over effect (ii). This is confirmed by the fact that λ is very close between three ($3d_{x^2-y^2}$-$3d_{z^2}$-4*s*) and two($3d_{x^2-y^2}$-$3d_{z^2}$, 4*s* effectively included) orbital models having the same Fermi surface shape[17,18].

Now, we can unify the present results into a simple picture of "orbital distillation". While the $3d_{x^2-y^2}$ orbital is supposed to compose a nearly square shape Fermi surface in itself, the Fermi surface is bent by the 4*s* orbital in the actual materials. Thus a strategy for having higher $T_c$ is to reduce the effects of $d_{z^2}$ and 4*s* orbitals *simultaneously* (as symbolized by thick arrows on the Fig.1 (a)), and we propose this "orbital distillation" as a key to optimize $T_c$.

4. Conclusion

In summary, we have studied the $3d_{x^2-y^2}$-$3d_{z^2}$-4*s* three-orbital model for $HgBa_2CuO_4$ derived from first principles. Applying FLEX approximation to this model, we have shown that the increase in the level offset $\Delta E_s$ enhances $T_c$ by suppressing the warping of the Fermi surface, and that the reduction of the inter-orbital hopping $t_s$ works in a similar way. Both results can be summarized in a single picture in which the 4*s* orbital makes the Fermi surface more warped and hence suppresses $T_c$. This result indicates the same tendency as those calculations adopting the single-orbital Hubbard Hamiltonian tuning $t_2$ and $t_3$ in the $3d_{x^2-y^2}$ orbital sector because the 4*s* orbital can be integrated out before the many body analysis. Still, we stress that the present study has given access to $T_c$-controlling parameters that are *more directly connected to the lattice structure, orbital symmetry of the constituent element*. It may be difficult to increase $\Delta E_s$ to greater extent in actual materials, but we believe there remains a possibility of reducing the inter-orbital hopping $t_s$. If such an orbital distillation is realized in some materials other than the cuprates, thereby preserving the "favourable conditions" enjoyed by the cuprates (near half-filling, $U/8t\sim 1$, square lattice), then there is a possibility that $T_c$ may be optimized even further.

**Acknowledgements**

The numerical calculations were performed at the Supercomputer Center, ISSP, University of Tokyo. This study has been supported by Grants-in-Aid for Scientific Research from JSPS (Grants No. 23340095, R.A.; No. 23009446, H.S.; No. 21008306, H.U.; No. 22340093, K.K. and H.A.). R.A. acknowledges financial support from JST-PRESTO. D.J.S. acknowledges support from the Center for Nanophase Material Science at Oak Ridge National Laboratory.

**References**

[1] J.D. Jorgensen, D.G. Hinks, O.Chmaissem, D.N. Argyiou, J.F. Mitchell and B. Dabrowski, in Lecture Notes in Physics, 475 (1996) p.1; J.D Jorgensen, D.G. Hinks, O.Chmaissem, D.N. Argyiou, J.F. Mitchell and B. Dabrowski, in: J.Klamut, B.W.Veal, B.M. Dabrowski, P.W. Klamut and M.Kazimierski (Eds.), Recent Developments in High Temperature Superconductivity, Springer, 1995, pp.1-15.
[2] A. Bianconi, G. Bianconi, S. Caprara, D. Di Castro, H. Oyanagi and N. L. Saini, J. Phys.: Condens. Matter 12 (2000) 10655; A. Bianconi, S. Agrestini, G. Bianconi, D. Di Castro and N. L. Saini, J. Alloys Compd. 537 (2001) 317; N. Poccia, A. Ricci and A. Bianconi, Adv. Condens. Matter Phys. 2010 (2010) 261849.
[3] H. Eisaki, N. Kaneko, D. L. Feng, A. Damascelli, P. K. Mang, K. M. Shen, Z.-X. Shen and M. Greven, Phys. Rev. B 69 (2004) 064512.
[4] E. Pavarini, I. Dasgupta, T. Saha-Dasgupta, O. Jepsen and O. K. Andersen, Phys. Rev. Lett. 87 (2001) 047003.
[5] K. Tanaka, T. Yoshida, A. Fujimori, D. H. Lu, Z.-X. Shen, X.-J. Zhou, H. Eisaki, Z. Hussain, S. Uchida, Y. Aiura, K. Ono, T. Sugaya, T. Mizuno and I. Terasaki, Phys. Rev. B 70 (2004) 092503.
[6] For a review, see D.J. Scalapino in: J.R. Schrieffer and J.S. Brooks (Eds.), *Handbook of High Temperature Superconductivity*, Springer, New York, 2007, Chapter 13,.
[7] K. Shiraishi, A. Oshiyama, N. Shima, T. Nakayama and H. Kamimura, Solid State Commun. 66 (1988) 629.
[8] H. Kamimura and M. Eto, J. Phys. Soc. Jpn. 59 (1990) 3053; M. Eto and H. Kamimura, J. Phys. Soc. Jpn. 60 (1991) 2311.
[9] A.J. Freeman and J. Yu, Physica B 150 (1988) 50.
[10] S. Maekawa, J. Inoue and T. Tohyama, in: Y. Iye and H. Yasuoka (Eds.), *The Physics and Chemistry of Oxide Superconductors*, Springer, Berlin, Vol. 60, 1992, pp. 105-115.
[11] O.K. Andersen, A.I. Liechtenstein, O. Jepsen and F. Paulsen, J. Phys. Chem. Solids 56 (1995) 1573.
[12] L.F. Feiner, J.H. Jefferson and R. Raimondi, Phys. Rev. Lett. 76 (1996) 4939.
[13] E. Pavarini, I. Dasgupta, T. Saha-Dasgupta, O. Jepsen and O. K. Andersen, Phys. Rev. Lett. 87 (2001) 047003.
[14] C. Weber, K. Haule and G. Kotliar, Phys. Rev. B 82 (2010) 125107.
[15] C. Weber, C. -H. Yee, K. Haule and G. Kotliar, arXiv:1108.3028.
[16] T. Takimoto, T. Hotta and K. Ueda, Phys. Rev. B 69 (2004) 104504.
[17] H. Sakakibara, H. Usui, K. Kuroki, R. Arita and H. Aoki, Phys. Rev. Lett. 105 (2010) 057003.
[18] H. Sakakibara, H. Usui, K. Kuroki, R. Arita and H. Aoki, Phys. Rev. B 85 (2012) 064501.
[19] X. Wang, H.T. Dang and A. J.Millis, Phys. Rev. B 84 (2011) 014530.
[20] L. Hozoi, L. Siurakshina, P. Fulde and J. van den Brink, Sci. Rep. 1 (2011) 65.
[21] S. Uebelacker and C. Honerkamp, Phys. Rev. B 85 (2012) 155122.
[22] M. Mori, G. Khaliullin, T. Tohyama and S. Maekawa, Phys. Rev. Lett. 101 (2008) 247003
[23] A.-K. Klehe, A. K. Gangopadhyay, J. Diederichs and J. S. Schilling, Physica C 213 (1993) 266; 223 (1994) 121.
[24] L. Gao, Y. Y. Xue, F. Chen, Q. Xiong, R. L. Meng, D. Ramirez, C. W. Chu, J.H Eggert and H.K. Mao, Phys. Rev. B 50 (1994) 4260.
[25] H. Sakakibara, K. Suzuki, H. Usui, K. Kuroki, R. Arita, D.J. Scalapino and Hideo Aoki, Phys. Rev. B 86 (2012) 134520.
[26] P. Blaha, K. Schwarz, G.K.H. Madsen, D. Kvasnicka and J. Luitz, Wien2k: An Augmented Plane Wave + Local Orbitals Program for Calculating Crystal Properties (Vienna University of Technology, Wien, 2001).
[27] F. Birch, Phys. Rev. B 71 (1947) 809.
[28] J.L. Wagner, P.G. Radaelli, D.G. Hinks, J.D. Jorgensen, J.F. Mitchell, B. Dabrowski, G.S. Knapp, M.A. Beno, Physica C 210 (1993) 447.
[29] J. Kunes, R. Arita, P. Wissgott, A. Toschi, H. Ikeda and K. Held, Comp. Phys. Commun. 181 (2010) 1888.
[30] N. Marzari and D. Vanderbilt, Phys. Rev. B 56 (1997) 12847; I. Souza, N. Marzari and D. Vanderbilt, Phys. Rev. B 65 (2001) 035109. The Wannier functions are generated by the code developed by A. A. Mostofi, J. R. Yates, N. Marzari, I. Souza and D. Vanderbilt, http://www.wannier.org/.
[31] N.E. Bickers, D.J. Scalapino and S.R. White, Phys. Rev. Lett. 62 (1989) 961.
[32] T. Dahm and L. Tewordt, Phys. Rev. Lett. 74 (1995) 793.
[33] K. Yada and H. Kontani, J. Phys. Soc. Jpn. 74 (2005) 2161.